\documentclass[useAMS,usenatbib]{mn2e}

\usepackage[normalem]{ulem} % to cross out part of the text

\usepackage{graphicx}
\usepackage{hyperref}
\usepackage{color,epsfig}
\usepackage[utf8]{inputenc}
\usepackage{verbatim}
\usepackage{lmodern}
\usepackage{amssymb}

\newcommand{\be}{\begin{equation}}
\newcommand{\ee}{\end{equation}}

\newcommand*\diff{\mathop{}\!\mathrm{d}}

\newcommand*{\eqref}[1]{(\ref{#1})}

\newcommand{\vect}[1]{\mathbf{#1}}

\def\msun{{\,{\rm M}_\odot}}
\def\rsun{{\,{\rm R}_\odot}}
\def\gcm3{\, \rm g \, cm^{-3}}
\def\kms{\, \rm km \, s^{-1}}
\def\pc{\, \rm pc}
\def\yr{\, \rm yr}

\def\rt{R_{\rm t}}

\def\mh{M_{\rm h}}

\def\mstar{M_{\star}}
\def\rstar{R_{\star}}

\def\tmin{t_{\rm min}}

\def\mdot{\dot{M}}

\def\rhozmw{\rho_{\rm 0,MW}}

\def\rzmw{R_{\rm 0,MW}}

\def\msun{M_{\odot}}

\def\me{m_{\rm e}}

\def\rhoe{\rho_{\rm e}}
\def\rhog{\rho_{\rm g}}

\def\rs{R_{\rm s}}

\def\ate{a_{\rm t,e}}
\def\age{a_{\rm g,e}}
\def\are{a_{\rm r,e}}
\def\ame{a_{\rm m,e}}
\def\ape{a_{\rm p,e}}

\def\fkhe{f_{\rm KH,e}}
\def\fkhp{f_{\rm KH,peak}}
\def\chiswa{\chi_{\rm swa}}
\def\chikh{\chi_{\rm KH}}
\def\de{\Delta \epsilon}
\def\rhostar{\rho_{\star}}

\def\frame{f_{\rm ram,e}}

\def\sgra{$\rm Sgr \, A^{*}$}

\def\le{l_{\rm e}}
\def\he{h_{\rm e}}
\def\me{m_{\rm e}}
\def\thetae{\theta_{\rm e}}
\def\ene{\epsilon_{\rm e}}
\def\dene{\delta \epsilon_{\rm e}}
\def\te{T_{\rm e}}
\def\dte{\delta T_{\rm e}}
\def\pe{P_{\rm e}}
\def\re{R_{\rm e}}
\def\ke{k_{\rm e}}
\def\ae{a_{\rm e}}
\def\ve{v_{\rm e}}
\def\tkhe{\tau_{\rm KH,e}}

\title[Bad prospects for giant stars' disruption]{Bad prospects for the detection of giant stars' tidal disruption: effect of the ambient medium on bound debris}
\author[Clément Bonnerot, Elena M. Rossi and Giuseppe Lodato]{Clément Bonnerot$^{1}$\thanks{E-mail: bonnerot@strw.leidenuniv.nl}, Elena M. Rossi$^{1}$ and Giuseppe Lodato$^{2}$\\
$^{1}$Leiden Observatory, Leiden University, PO Box 9513, 2300 RA, Leiden, the Netherlands\\
$^{2}$Dipartimento
 di Fisica, Università Degli Studi di Milano, Via Celoria, 16, Milano, 20133, Italy
}
\begin{document}

\date{ Accepted ?. Received ?; in original form ?}

\pagerange{\pageref{firstpage}--\pageref{lastpage}} \pubyear{2014}

\maketitle

\label{firstpage}

\begin{abstract}
Most massive galaxies are thought to contain a supermassive black hole in their centre surrounded by a tenuous gas environment, leading to no significant emission. In these quiescent galaxies, tidal disruption events represent a powerful detection method for the central black hole. Following the disruption, the stellar debris evolves into an elongated gas stream, which partly falls back towards the disruption site and accretes onto the black hole producing a luminous flare. Using an analytical treatment, we investigate the interaction between the debris stream and the gas environment of quiescent galaxies. Although we find dynamical effects to be negligible, we demonstrate that Kelvin–Helmholtz instability can lead to the dissolution of the stream into the ambient medium before it reaches the black hole, likely dimming the associated flare. This result is robust against the presence of a typical stellar magnetic field and fast cooling within the stream. Furthermore, we find this effect to be enhanced for disruptions involving more massive black holes and/or giant stars. Consequently, although disruptions of evolved stars have been proposed as a useful probe of black holes with masses $\gtrsim 10^8 \msun$, we argue that the associated flares are likely less luminous than expected.
\end{abstract}


\section{Introduction}

Tidal disruption events (TDEs) occur when a star is scattered into a plunging orbit that brings it so close to a supermassive black hole (SMBH) that it is torn apart by strong tidal forces \citep{frank1976,rees1988}. During the disruption, the stellar elements are forced into different trajectories, which causes the debris to subsequently evolve into an elongated gas stream. Half of the debris within this stream is bound to the black hole while the other half is unbound. After a revolution around the black hole, the bound debris returns to the disruption site and forms an accretion disc \citep{shiokawa2015,hayasaki2013, hayasaki2015,bonnerot2015}, from which a powerful flare can be emitted \citetext{\citealt{komossa2004}, \citealt{gezari2012}, see \citealt{komossa2015} for a recent review}. This flare represents a unique probe to detect SMBHs in the centres of otherwise quiescent galaxies. Through this signal, it is also in principle possible to put constraints on the black hole properties as well as to investigate the physics of accretion and relativistic jets around these objects.

The debris evolution within the stream from disruption to its return to pericentre has been the focus of several studies, both numerical and analytical. While the debris follows close to ballistic orbits, the transverse structure of the stream is set by the equilibrium between the different forces acting in this direction. During most of its evolution, internal pressure is balanced by self-gravity, which causes the stream to maintain a narrow profile \citep{kochanek1994,ramirez-ruiz2009,guillochon2014}. However, a recent simulation shows that internal pressure inside the stream may be unable to prevent the fragmentation of the debris into self-gravitating clumps, which can form a few years after disruption \citep{coughlin2015,coughlin2015_2}.

Although it is not associated to substantial emission, a gas component is present around SMBHs in the centre of quiescent galaxies. It is commonly assumed to originate from stellar winds released by massive stars surrounding the black hole \citep{quataert2004,cuadra2006,generozov2015}. The impact of this gaseous environment on the stream evolution has so far been largely ignored, owing to a large density contrast between the two components. In a recent study, \citet{guillochon2015_3} find that it can affect the trajectories of the unbound debris, resulting in its deceleration on parsec scales. Other authors looked into the influence on the bound part of the stream but in specific contexts, such as a possible origin for the G2 cloud \citep{guillochon2014} and the interaction with a fossil accretion disc \citep{kelley2014}.

In this paper, we investigate the influence of the ambient gas on the bound debris in a general way. Although dynamical effects are negligible, we demonstrate that hydrodynamical instabilities can lead to the dissolution of a significant part of this debris into the gaseous environment before it returns to pericentre. In this situation, we argue that the associated TDE would be significantly dimmer than expected. This effect is enhanced when the disruption involves a giant star and/or a more massive black hole. As a result, TDEs involving black holes of mass $\gtrsim 10^8 \msun$ could be difficult to detect. While main sequence stars are swallowed whole by such black holes leading to no substantial emission \citep{macleod2012}, disruptions of giant stars could be just as dim owing to the dissolution of the debris into the ambient medium.

This paper is organized as follows. Sections \ref{gaseous} and \ref{stream} present the models used for the SMBH gaseous environment and the debris stream respectively. Section \ref{interactions} investigates the interaction between these two components through both ram pressure and hydrodynamical instabilities. In Section \ref{fraction}, we determine the impact on the detectability of TDEs. Our concluding remarks are found in Section \ref{conclusion}.

\section{Gaseous environment model}
\label{gaseous}

In quiescent galaxies, black holes are surrounded by accretion flows, whose gas is mostly supplied by stellar winds from massive stars. The density distribution within this flow is given by the interplay between their hydrodynamics and the efficiency of the supply mechanism.

The Milky Way is the best example of a quiescent galaxy. It harbours \sgra, a central black hole of mass $4.3 \times 10^6 \msun$, surrounded by a gas environment well studied both theoretically and observationally. Analytical models of stellar winds sources find a density profile in the inner region of the flow decreasing as $R^{-1}$ \citep{quataert2004,generozov2015}, a result consistent with numerical simulations \citep{cuadra2006}.

Based on this example, we adopt a simple gas density profile for the inner region of quiescent galaxies, given by\footnote{In our galaxy, a density profile scaling as $R^{-1/2}$ may be more consistent with observations of the inner accretion flow \citep{wang2013}. In other quiescent galaxies, this slope can be derived from observations of TDEs featuring outflows, where it is found to be steeper, decreasing as $R^{-5/2}$ \citep{alexander2015} or $R^{-3/2}$ \citep{berger2012}. However, this could be caused by the propagation of the outflow into a previously evacuated funnel.}
\be
\rhog(R) =  \rho_0 \left( \frac{R}{R_0} \right)^{-1},
\label{gasdensity}
\ee
For the Milky Way, the normalization is inferred from Chandra X-ray observations at the Bondi radius, which find a density $\rhozmw=2.2 \times 10^{-22} \gcm3$ at $\rzmw=0.04 \pc$. 

For galaxies hosting SMBHs of different masses, this profile is scaled using the black hole radius of influence
\be
R_{\rm inf}=\frac{G \mh}{\sigma^2} \simeq 3 \pc \left(\frac{\mh}{4.3 \times 10^6 \msun} \right)^{7/15},
\ee
where $\mh$ is the black hole mass, $\sigma$ is the velocity dispersion of stars in the bulge and the second equality uses the $\mh - \sigma $ relation $\mh= 2 \times 10^8 (\sigma / 200 \kms)^{15/4} \msun$ \citep{gebhardt2000}\footnote{The $\mh-\sigma$ relation can be steeper than this. However, our results are essentially unchanged when using a steeper $\mh \propto \sigma^{5.3}$ relation \citep{mcconnell2011}.}. The normalization radius is then obtained from
\be
R_0=  \left(\frac{\mh}{4.3 \times 10^6 \msun} \right)^{7/15} \rzmw.
\label{rzero}
\ee
The normalization density is computed by assuming spherical accretion at a velocity $v \propto v_{\rm ff} \propto \mh^{1/2} R^{-1/2}$, where $v_{\rm ff}$ is the free-fall velocity. It leads to an accretion rate $\mdot \propto R^2_0  \rho_0 v(R_0) \propto \rho_0\mh^{6/5} $ using equation \eqref{rzero}. The gas is supplied to the accretion flow by stellar winds from stars within the black hole sphere of influence. As the mass of stars is similar to that of the black hole within this distance, $\mdot \propto \mh$. This yields
\be
\rho_0= \eta \left( \frac{\mh}{4.3 \times 10^6 \msun}\right)^{-1/5}\rhozmw,
\label{rhozero}
\ee
where $\eta$ is a parameter, equal to 1 for the Milky Way. In the following, it is varied up to 1000 to investigate galaxies with denser gas environments. This simple scaling of the gas density profile has also been used by \citet{rimoldi2015}. It leads to a similar dependence on $\mh$ as found from a more detailed treatment \citep{generozov2015}.

\begin{figure}
\epsfig{width=0.47\textwidth, file=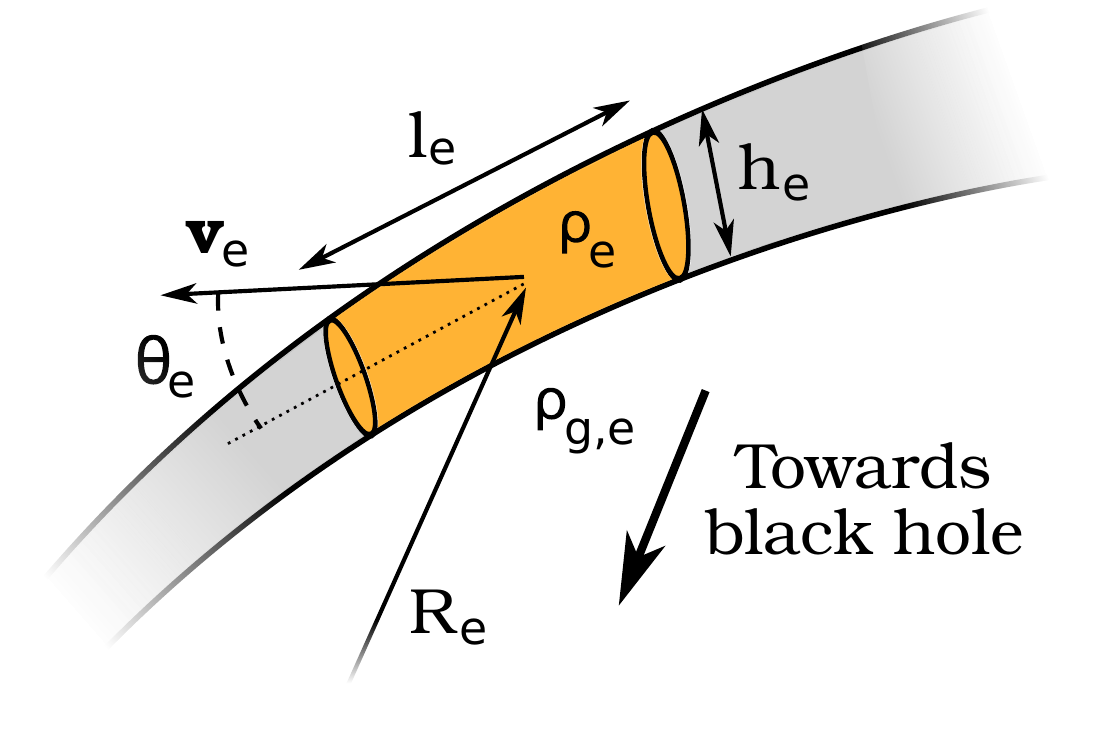}
\caption{Sketch of a portion of debris stream with an element shown in orange. The element has a cylindrical geometry, with length $\le$ and width $\he$. Its density $\rhoe$ is obtained from equation \eqref{rhoe_cyl} knowing its mass. At a distance $\re$ from the black hole, it moves through a gaseous environment of density $\rho_{\rm g,e} \equiv \rhog(\re)$ with a velocity $\vect{v_{\rm e}}$, inclined with respect to its longitudinal axis by an angle $\thetae$.}
\label{fig1}
\end{figure}

\section{Tidal stream model}
\label{stream}

The disruption of a star of mass $\mstar$ and radius $\rstar$ occurs when it reaches the tidal radius $\rt=\rstar (\mh/\mstar)^{1/3}$. The resulting debris evolves into an elongated stream owing to an orbital energy spread $\de=G\mh \rstar /R^2_{\rm t}$, acquired during the disruption. In this work, we only focus on the bound debris, with orbital energies $\epsilon$ from $-\de$ to 0 and periods $T$ between $\tmin=2 \pi G \mh (2 \de)^{-3/2}$ and $+\infty$.

To model the stream of bound debris, we divide it into cylindrical elements, an example of which is sketched in Fig. \ref{fig1}. In the following, the variables associated to a particular element are indicated by the subscript ``$\rm e$'' to differentiate them from those associated to the debris.

An element of period $\te$ contains debris whose periods satisfy $\te-\dte<T<\te+\dte$. Equivalently, it has an average orbital energy $\ene = -(1/2)(2 \pi G \mh /\te)^{2/3}$ and contains debris with orbital energies in the range $\ene- \dene<\epsilon<\ene+\dene$. To ensure that each element is composed of debris with similar periods, we set $\dte = 10^{-2} \, \tmin \ll \te$. 

Following the disruption, each component of the stream is assumed to follow Keplerian orbits with the same pericentre $\rt$ but different orbital energies $\epsilon$. The position $\vect{x_{\rm e}}$ and velocity $\vect{v_{\rm e}}$ of an element are identified with those of the debris with orbital energy $\ene$.

Owing to its cylindrical geometry, the density of an element is obtained by
\be
\rhoe = \frac{\me}{\pi h^2_{\rm e} \le},
\label{rhoe_cyl}
\ee
where $\me$, $\he$ and $\le$ denote the mass, width and length of the element respectively. We explain how these quantities are computed in the remaining of this section.

Knowing the separation $\vect{\delta x_{\rm e}}$ of its two extremities, the length of an element is obtained by $\le=|\vect{\delta x_{\rm e}}|$. Its velocity $\vect{v_{\rm e}}$ is inclined with respect to its longitudinal direction by an angle $\thetae$ obtained by $\cos \thetae = \vect{v_{\rm e}} \cdot \vect{\delta x_{\rm e}}/ (|\vect{v_{\rm e}}| |\vect{\delta x_{\rm e}}|)$.

The mass $\me$ of an element is obtained from
\be
\me=\int_{\ene - \dene}^{\ene + \dene} \diff M \simeq 2 \frac{\diff M}{\diff \epsilon}\Big|_{\ene}  \dene.
\ee
where $\diff M / \diff \epsilon$ is the debris orbital energy distribution. The latter is computed using the analytical model developed by \citet{lodato2009}, which assumes that the debris energy is given by its depth within the black hole potential when the star is disrupted. This yields
\be
\frac{\diff M}{\diff \epsilon} = \frac{\rstar}{\de} \int_{\Delta r}^{\rstar} 2 \pi \rhostar(r) r \diff r,
\ee
where $\rhostar$ is the density inside the star and $\Delta r = (\epsilon/\de) \rstar$. This allows to compute the fallback rate of the debris to pericentre, given by
\be
\dot{M}_{\rm fb} = \frac{\diff M}{\diff \epsilon} \frac{\diff \epsilon}{\diff T} = \frac{(2 \pi G \mh)^{2/3}}{3} \frac{\diff M}{\diff \epsilon} T^{-5/3},
\label{mdotfb}
\ee
where the relation $T=2 \pi G \mh (-2 \epsilon)^{-3/2}$ is used in the second equality.

\begin{figure}
\epsfig{width=0.47\textwidth, file=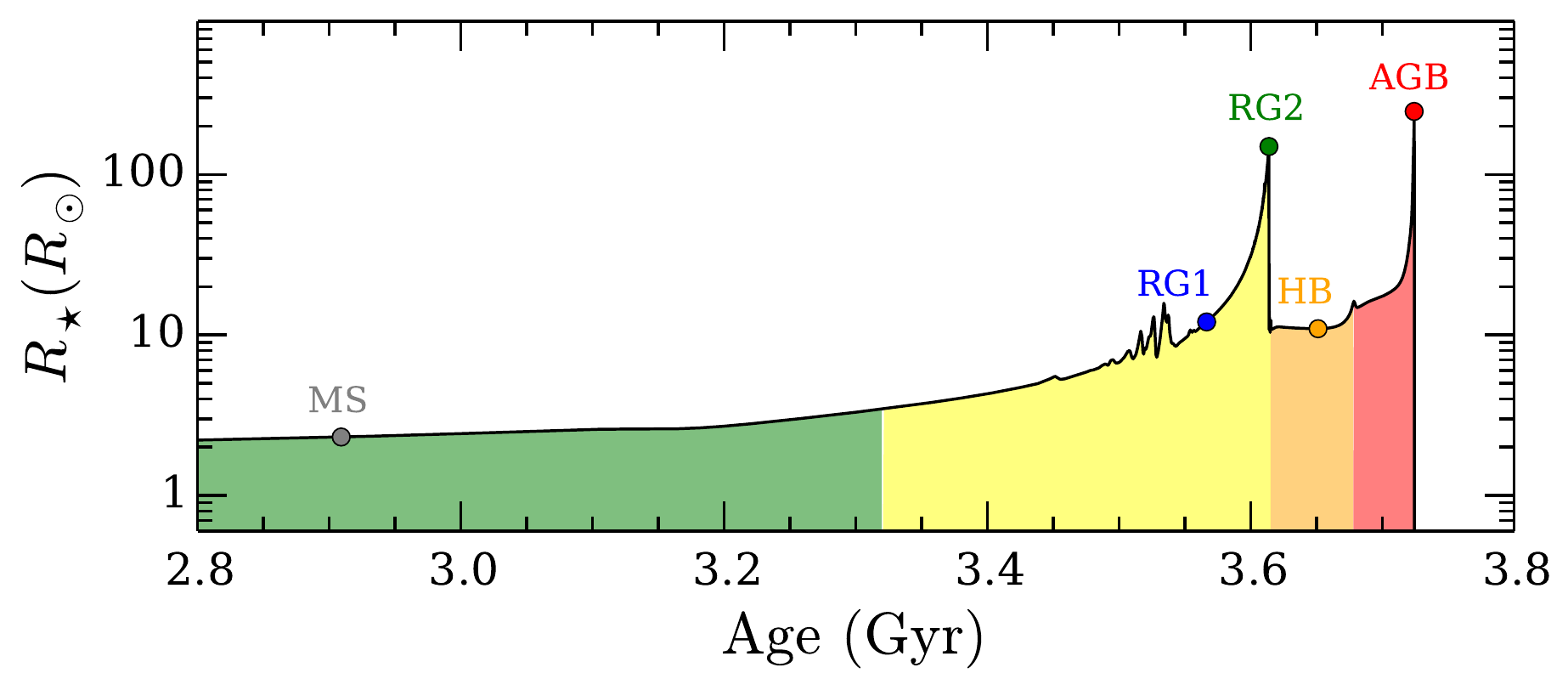}
\caption{Evolution of the radius of a 1.4 $\msun$ star. The main evolutionary phases are indicated by filled regions: main sequence (green), red giant (yellow), horizontal branch (orange) and asymptotic giant branch (red). The coloured points correspond to the five stellar density profiles considered.}
\label{fig2}
\end{figure}

Based on the work by \citet{macleod2012}, different density profiles are considered corresponding to the evolution of a 1.4 $\msun$ star. They are obtained from a detailed simulation of the star using the stellar evolution code MESA \citep{paxton2011}. The evolution of the stellar radius is shown in Fig. \ref{fig2}, with the main phases of evolution indicated by filled areas and the five stellar density profiles considered later in the paper shown with coloured points. In the main sequence phase (green area), one profile is considered (MS). Two profiles are chosen in the red giant phase (yellow area): when the star is ascending the red giant branch (RG1) and when it reached the tip of this branch (RG2). For the horizontal branch (orange area) and the asymptotic giant branch (red area) phases, two profiles are selected (HB and AGB).

The width $\he$ is obtained by assuming hydrostatic equilibrium in the stream transverse direction. While pressure tends to expand the stream, the tidal force from the black hole and the stream self-gravity oppose this expansion. Note that the tidal force acts inwards since the stream transverse direction is close to that orthogonal to the direction of the black hole. Hydrostatic equilibrium thus reduces to
\be
\ape=\ate+\age
\label{widtheq}
\ee
where $\ape = \nabla \pe / \rhoe \simeq \pe / (\rhoe \he)$ is the pressure acceleration, $\ate \simeq G \mh \he / \re^3 $ is the tidal acceleration and $\age\simeq G \me/ (\he\le)$ is the self-gravity acceleration within the stream, $\re=|\vect{x_{\rm e}}|$ being the distance from the black hole and $\pe$ the pressure in the stream. For the pressure, we assume an adiabatic evolution with $\pe=K \rhoe^{\gamma}$ where $\gamma=5/3$. Although the adiabatic constant $K$ should a priori be different for different elements, we adopt a single value averaged over the volume of the star. This is legitimate as the value of $K$ within the star varies only by a factor of a few around this average. The width $\he$ is obtained by solving equation \eqref{widtheq}, making use of equation \eqref{rhoe_cyl}. For illustration, in the two limiting cases $\age \gg \ate$ and $\age \ll \ate$, it scales as $\he \propto (\me/\le)^{-1/4}$ and $\he \propto (\re^3/ \mh)^{3/10} (\me/\le)^{-1/5}$ respectively.

\section{Tidal stream - ambient medium interactions}
\label{interactions}

\subsection{Hydrodynamical instabilities}

As the stream moves through the ambient medium, it is subject to the Kelvin–Helmholtz (K-H) instability. In this section, we evaluate the effect of this instability on each stream element.

Taking a conservative approach, we only consider the second half of each element orbit, i.e. after apocentre passage. This approach is motivated by the fact that an element reaches its lowest density in this part of the orbit and is therefore more easily affected by its interaction with the ambient medium. In this portion of the orbit, an element falls almost radially from apocentre to pericentre. In this configuration, the K-H instability develops on a given stream element for wavenumbers $\ke$ which obey the inequality \citep[p.138]{clarke2007}
\be
\ae  < \frac{\rhoe \, \rho_{\rm g,e}}{\rhoe^2-\rho^2_{\rm g,e}} \, \ke \, v^2_{\rm rel,e} 
\label{condgen}
\ee
where $\ae$ is the inwards acceleration of the element in the transverse direction, $\rho_{\rm g,e}\equiv \rhog(\re)$ is the density of gas at the position of the element and $v_{\rm rel,e}$ is the relative velocity between the element and the background gas. Although modes with large $\ke$ have fast growth rates, they are also the least disruptive as the associated instability saturates at an amplitude $\sim 1/\ke$. We therefore consider a wavenumber $\ke=1/\he$ which has the slowest growth rate but is the most disruptive since it develops on the whole element width. The transverse acceleration $\ae$ has two inwards components. One is the self-gravity acceleration $\age\simeq G \me / (\he\le)$ and the other is the tidal acceleration $\ate \simeq G \mh \he / R^3_{\rm e}$. With $\ae=\ate+\age$, condition \eqref{condgen} reduces to 
\be
\ate+\age<\frac{\rho_{\rm g,e} v^2_{\rm e}}{\rhoe \he}\equiv \are,
\label{cond}
\ee
which uses $\rhoe \gg \rho_{\rm g,e}$. The relative velocity is computed by $v_{\rm rel,e} = \ve \cos \thetae \simeq \ve$ where $\ve = |\vect{v_{\rm e}}|$ is the velocity of the element. This uses the approximation $\thetae \ll 1$, which is satisfied along an element orbit, as soon as it leaves its apocentre. In addition, this value of $v_{\rm rel,e}$ assumes that the background gas is at rest. The possibility of a lower relative velocity caused by radially falling back ground gas has been explored and leads to no significant difference. The right-hand side of equation \eqref{cond} is called $\are$ as it is equivalent to a ram pressure acceleration.

If condition \eqref{cond} is satisfied, the K-H instability then grows on a timescale

\be
\tkhe=\left(\frac{\he}{\are-\ate-\age}\right)^{1/2},
\label{tkh}
\ee
for a given element. Otherwise, the instability does not develop and $\tkhe = +\infty$. The K-H instability has time to fully grow before the element reaches pericentre if
\be
\fkhe \equiv\int_{\te/2}^{\te} \frac{\rm dt}{\tkhe} >1,
\label{fkhcond}
\ee
where $\te/2$ and $\te$ are the times corresponding to the element apocentre and pericentre passages respectively. Condition \eqref{fkhcond} can be understood by omitting the temporal dependence of $\tkhe$. In this case, it reduces to $\tkhe < \te/2$ which clearly implies that the K-H instability has time to fully grow during the portion of orbit considered.

\begin{figure}
\epsfig{width=0.47\textwidth, file=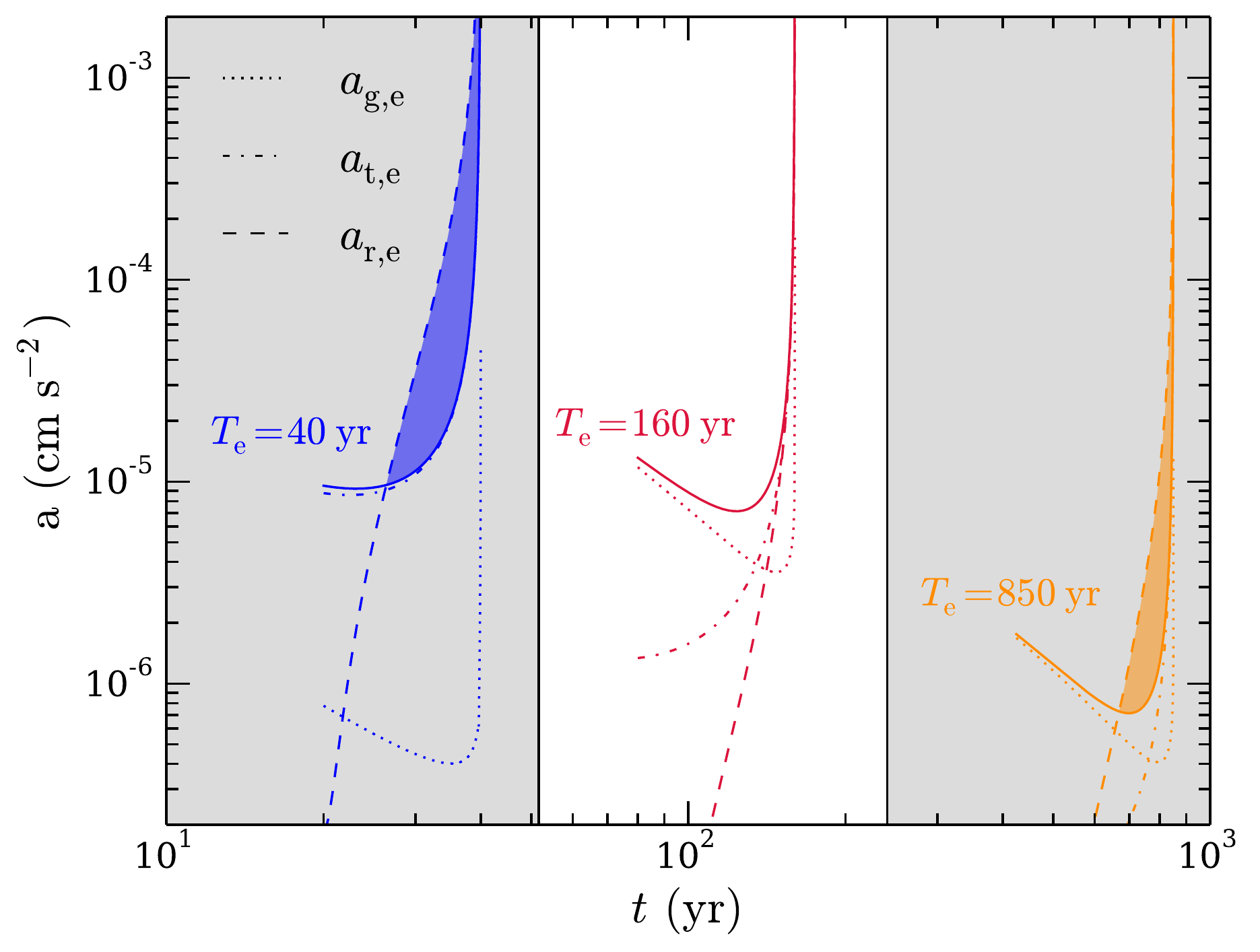}
\caption{Evolution of $\age$ (dotted lines), $\ate$ (dashed dotted lines), $\age+\ate$ (solid line) and $\are$ (dashed lines) for three elements of a stream produced by the tidal disruption of the star in the red giant phase (profile RG1) by a black hole of mass $\mh=10^8 \msun$ in a galaxy with $\eta = 5$. The elements have different periods $\te=40 \yr$ (blue lines), $\te=160 \yr$ (red lines) and $\te=850 \yr$ (yellow lines). For each element, the filled areas indicate the regions where $\age+\ate<\are$, that is where condition \eqref{cond} is satisfied. The grey areas indicate the range of periods of elements verifying condition \eqref{fkhcond}, for which the K-H instability has time to fully develop before they return to pericentre.}
\label{fig3.pdf}
\end{figure}

As an example, the evolution of $\age$ (dotted lines), $\ate$ (dashed dotted lines), the left-hand side of equation \eqref{cond} $\age+\ate$ (solid line) and its right-hand side $\are$ (dashed lines) is shown in Fig. \ref{fig3.pdf} for three different elements of a stream produced by the disruption of the star in the red giant phase (profile RG1) by a black hole of mass $\mh=10^8 \msun$ in a galaxy with $\eta=5$. These elements have periods $\te=40 \yr$ (blue lines), $\te=160 \yr$ (red lines) and $\te=850 \yr$ (yellow lines). For all elements, tidal acceleration dominates self-gravity acceleration ($\ate>\age$) in the final part of their orbit, when $\re<\he(\mh/\me)^{1/3}$. The zones where condition \eqref{cond} is true, are indicated by filled regions for each element. They only exist for the most bound (blue lines) and least bound (yellow lines) of the elements considered. For these two elements, condition \eqref{fkhcond} is also satisfied and the K-H instability therefore has time to fully develop before they return to pericentre. For the intermediate element (red lines), condition \eqref{cond} is never verified. This implies $\fkhe=0$ and condition \eqref{fkhcond} is therefore not satisfied either. The grey areas indicate the range of periods of all the stream elements that satisfy condition \eqref{fkhcond}. On these elements, we expect the K-H instability to fully grow over the course of their orbit.

Fig. \ref{fig3.pdf} indicates the range of periods of elements that satisfy condition \eqref{fkhcond}, but does not show the precise evolution of $\fkhe$ with $\te$. Actually, the transition between  $\fkhe=0$ and $\fkhe>1$ is very sharp. For elements that never satisfy condition \eqref{cond}, $\fkhe=0$. However, as soon as condition \eqref{cond} is met at some point along an element orbit, $\fkhe\gtrsim 1$, which implies that condition \eqref{fkhcond} is already marginally satisfied. This is because, in the final part of an element orbit where $\ate \gg \age$, the inequality $\are / \ate \gtrsim 2$ implies $\tkhe \lesssim (G \mh / R^3_{\rm e})^{-1/2}$, where the right-hand side is the infall time from $\re$ to pericentre. Omitting the time dependence of $\tkhe$, this translates to $\fkhe \gtrsim 1$.

The reason why only the most and least bound part of the stream are affected by the K-H instability can be understood by examining condition \eqref{cond} more in detail in the final part of each element orbit, where $\ate \gg \age$. Using $\ve \simeq (G \mh/\re)^{1/2}$, it reduces to
\be
\me/\le<\rho_{\rm g,e} R^2_{\rm e},
\label{simpcond}
\ee
that is a condition on the stream linear density. Note that this condition is also independent on the element width $\he$. Our results are therefore largely independent on the assumption of hydrostatic equilibrium made to compute this width in Section \ref{stream}. Furthermore, this means that physical mechanisms modifying the stream width, such as fast cooling of the debris, are unlikely to affect our results. One can clearly see that condition \eqref{simpcond}, and therefore condition \eqref{fkhcond}, is easily satisfied for the most bound part of the stream, which is less massive since it originates from the tenuous outer layer of the star. Although the least bound part of the stream contains more mass, it is stretched owing to different trajectories of neighbouring debris regions and condition \eqref{simpcond} is also satisfied. 

At this point, one can predict how the impact of the K-H instability depends on the other parameters, namely the black hole mass $\mh$, the evolutionary stage of the star and $\eta$, which relates to the ambient medium density via equation \eqref{rhozero}. Tidal disruptions by more massive black holes lead to more extended streams. Furthermore, the right-hand side of condition \eqref{simpcond} evaluated at $\rt$ scales as $\rho_{\rm g,e} R^2_{\rm e} \simeq \rhog(\rt) R^2_{\rm t} \propto \mh^{9/15}$, which increases with the black hole mass. We therefore anticipate condition \eqref{cond} to be more easily satisfied when $\mh$ is larger. The stream is therefore likely to be more sensitive to the K-H instability. This trend is also expected for disruptions of evolved stars as they also lead to more extended streams whose debris originates from a more tenuous outer layer. Finally, we anticipate the same tendency when $\eta$ is increased, that is for environments with higher gas density, since $\rho_{\rm g,e} R^2_{\rm e} \propto \eta$. These predictions will be verified explicitly in Section \ref{luminosity}.

\begin{figure}
\epsfig{width=0.47\textwidth, file=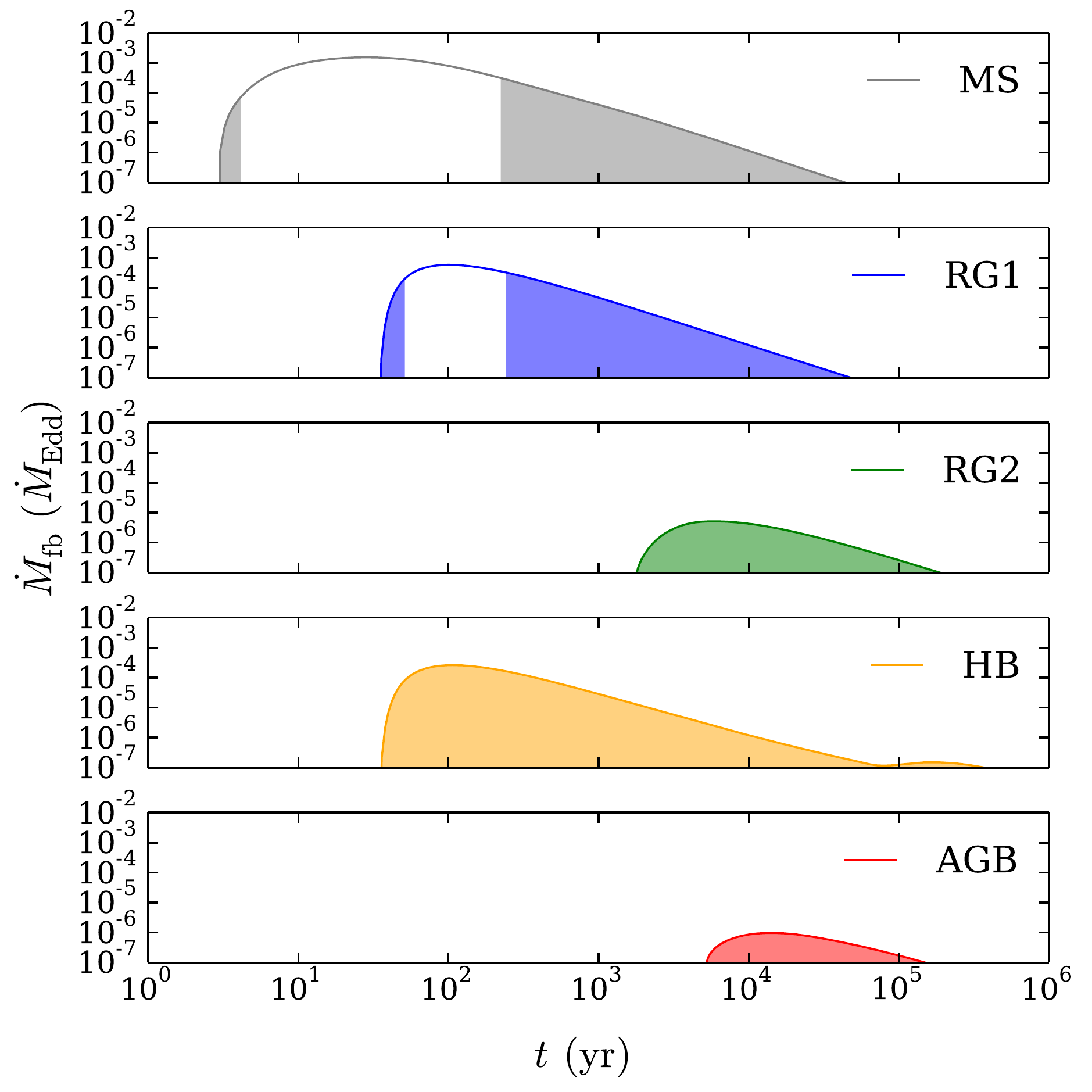}

\caption{Evolution of the debris mass fallback rate after a disruption with $\mh=10^8 \msun$ and $\eta=5$ for the five stellar density profiles considered: MS (grey line), RG1 (blue line), RG2 (green line), HB (yellow line) and AGB (red line). The filled areas correspond to the return times of debris satisfying condition \eqref{fkhcond}.}
\label{fig4}
\end{figure}

\begin{figure}
\epsfig{width=0.47\textwidth, file=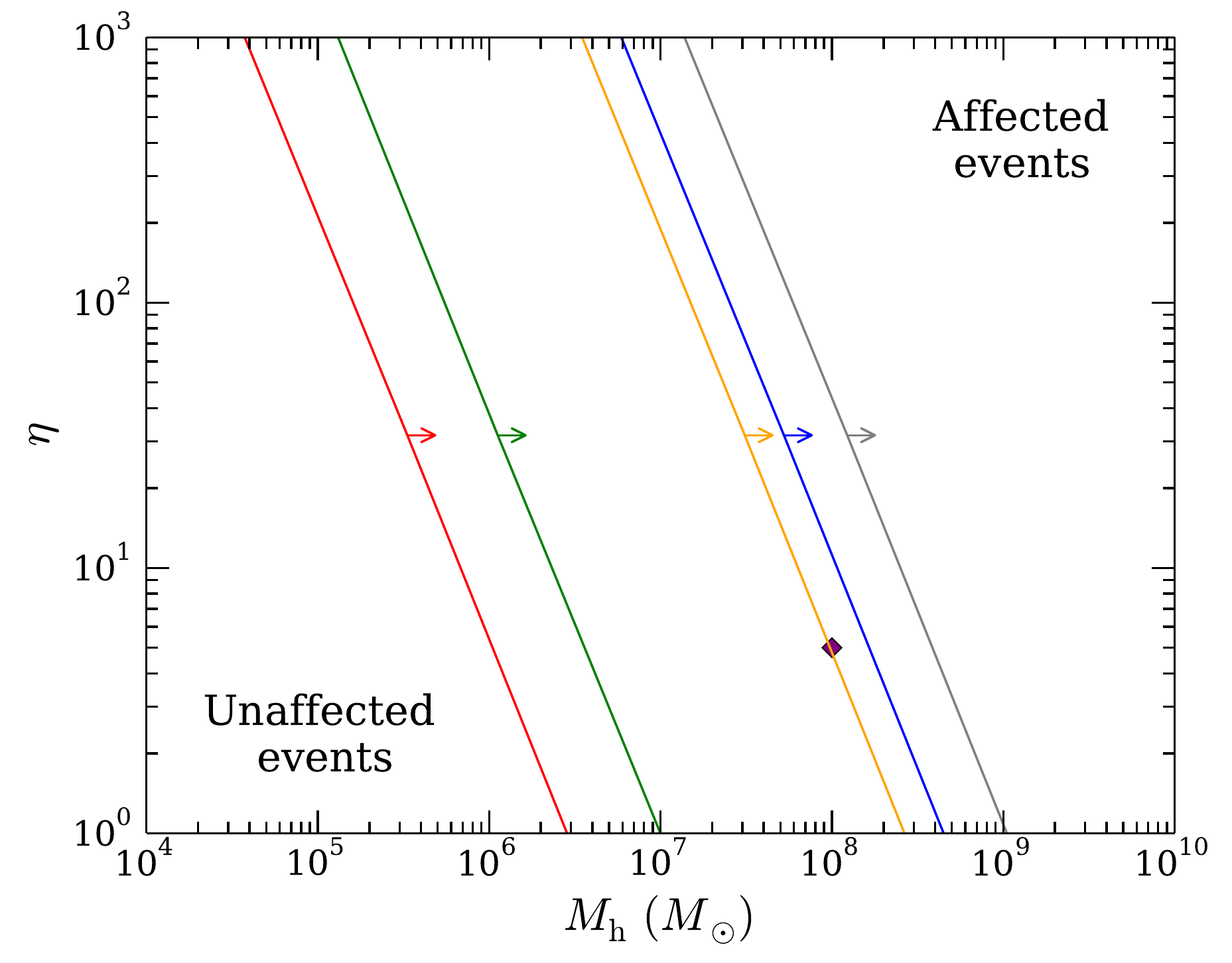}
\caption{$M_{\rm h}-\eta$ plane depicting the effect of the K-H instability on different disruption events. Each line corresponds to one of the stellar density profiles considered. The zone in the direction of the arrow corresponds to events affected by the K-H instability, for which $\fkhp>1$. The zone in the opposite direction corresponds to events for which $\fkhp<1$, unaffected by the K-H instability. The purple diamond shows the parameters corresponding to Fig. \ref{fig4}, $\mh=10^8 \msun$ and $\eta=5$.}
\label{fig5}
\end{figure}

\subsection{Ram pressure}

As a stream element sweeps up the ambient medium located on its trajectory, it loses momentum and decelerates. This deceleration affects significantly the trajectory of the element once it has swept a mass of ambient gas larger than its own mass. This is equivalent to
\be
\frame \equiv \frac{1}{\me} \int_{0}^{\te} \rhoe \, \ve \rm A_{\rm e} \, \diff t > 1,
\label{framcond}
\ee
where $A_{\rm e}=\he \le \sin \thetae$ is the element area sweeping gas from the ambient medium. As for the K-H instability, we find this condition to be satisfied both for the most and least bound part of the stream. However, $\frame<\fkhe$ in all cases explored, which means that the debris is affected by the K-H instability before their trajectories change due to ram pressure.

\subsection{Effect on flare luminosities}
\label{luminosity}

We now evaluate the impact of the K-H instability on the flare luminosities produced by the disruption of the star in different evolutionary stages and examine the dependence on the black hole mass $\mh$ and ambient gas density, through the parameter $\eta$.

Fig. \ref{fig4} shows the fallback rate, computed using equation \eqref{mdotfb}, of the debris produced by the disruption of the star by a black hole of mass $\mh=10^8\msun$ in a galaxy with $\eta=5$ for the five stellar density profiles considered. The filled areas indicate the times at which elements satisfying condition \eqref{fkhcond} return to pericentre. For these elements, the K-H instability has time to fully grow over the course of their orbit. For profiles MS and RG1, these zones exist only for the most and least bound debris, as in the example of Fig. \ref{fig3.pdf}. The debris whose return times correspond to the peak fallback rate are always outside this zone. Instead, for profiles RG2, HB and AGB, all the elements lie in the filled zone, even those returning to pericentre when the fallback rate peaks. It means that the K-H instability has time to fully grow in the whole stream. This confirms our expectation that streams produced by the disruption of evolved stars are more sensitive to the K-H instability.

So far, we have examined for which elements condition \eqref{fkhcond} is satisfied, that is for which debris the K-H instability fully develop before it reaches pericentre. As these instabilities involve the whole width of the stream, we infer that this debris subsequently dissolves into the ambient medium and does not return to pericentre.

Only the elements reaching pericentre intact can participate to the luminosity emitted from the event. Therefore, if all the stream dissolves into the ambient medium due to the K-H instability, the appearance of the event is likely to be affected, emitting a significantly lower luminosity. We take a conservative approach and state that an event is affected by this instability if even the element corresponding to the peak of the mass fallback rate dissolves into the background gas. According to our criterion, this requires condition \eqref{fkhcond} to be satisfied for this element, that is $\fkhp > 1$. Fig. \ref{fig5} shows the regions of the $\mh-\eta$ plane corresponding to events affected by the K-H instability. Each line is associated to one of the stellar density profiles considered. The zone in the direction of the arrow corresponds to affected events while the zone in the opposite direction corresponds to unaffected events. The example discussed above ($\mh = 10^8 \msun$ and $\eta = 5$), where events corresponding to profiles RG2, HB and AGB are affected, is indicated by a purple diamond. As predicted above, events involving more massive black holes or occurring in galactic nuclei with denser gaseous environment are more sensitive to the K-H instability.

\begin{figure*}
\epsfig{width=0.47\textwidth, file=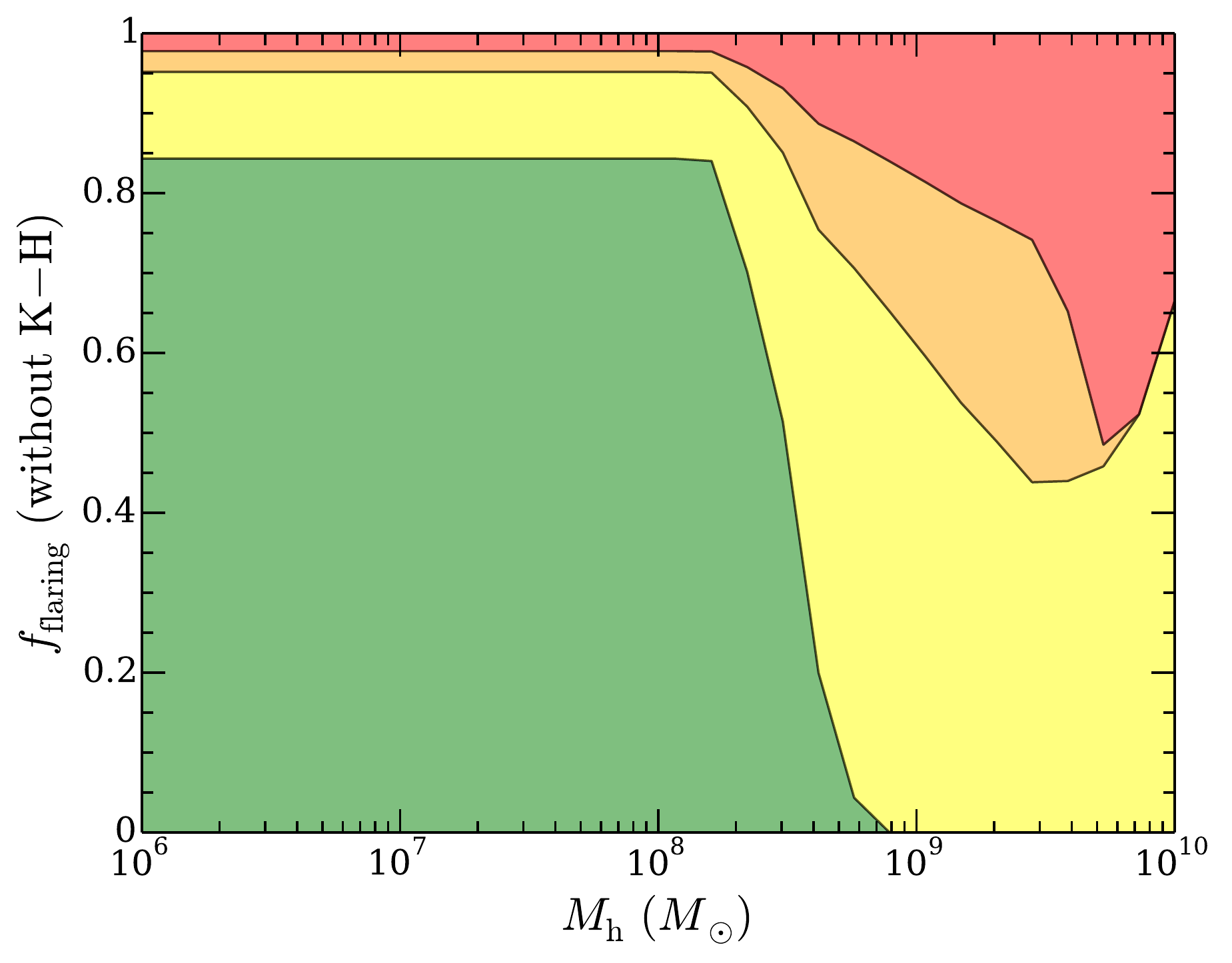}
\epsfig{width=0.47\textwidth, file=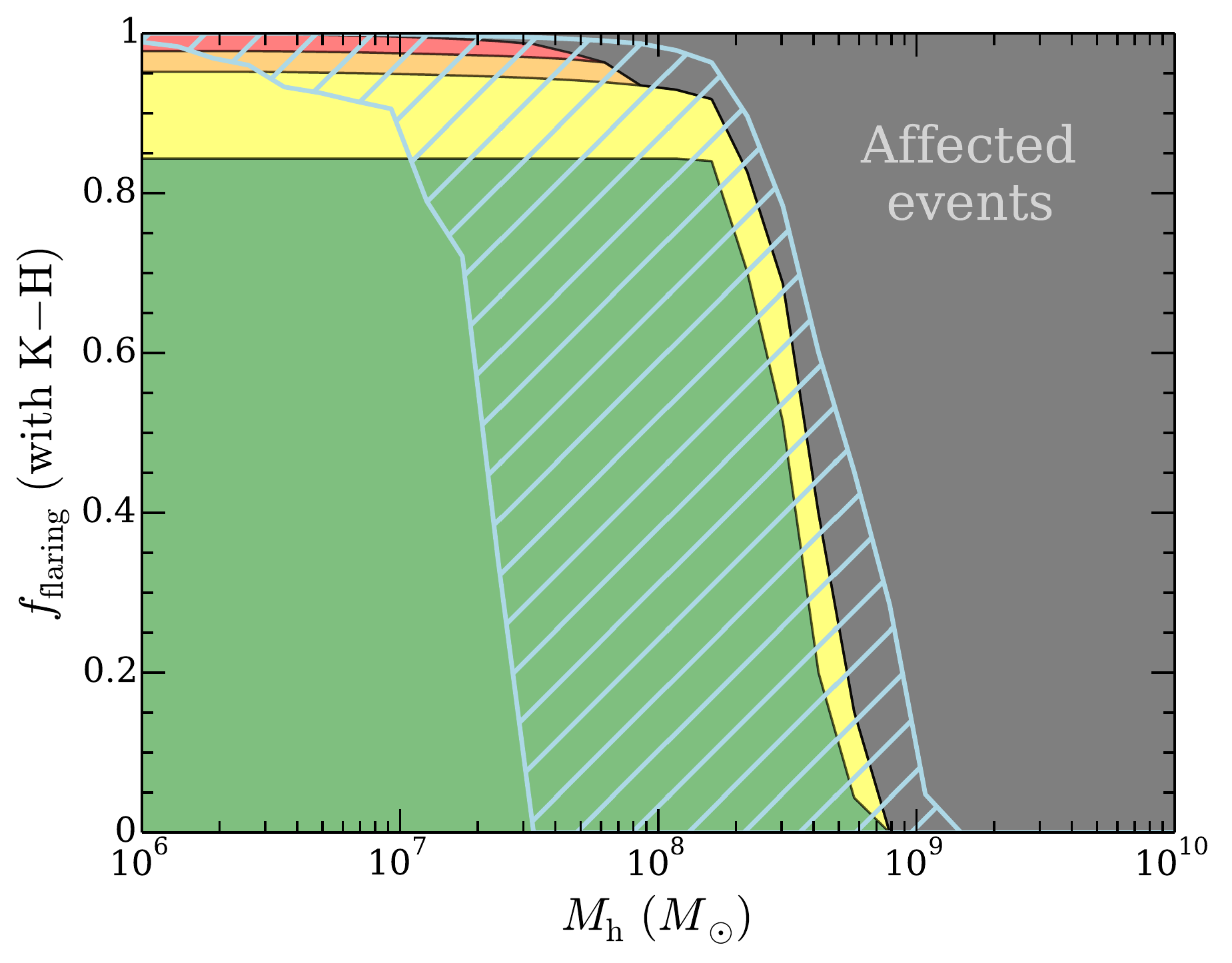}
\caption{Probability for a disruption event to occur in a given evolutionary stage as a function of $\mh$. The different coloured areas correspond to different phases in the evolution of the star: main sequence (green), red giant (yellow), horizontal branch (orange) and asymptotic giant branch (red). Only the right panel includes the effect of the K-H instability, with the grey area corresponding to affected events for a galaxy with $\eta=5$. The zone swept by the boundary of this area is shown by a blue hatched region for values of $\eta$ varying from 1 to 1000 from right to left.}
\label{fig6}
\end{figure*}

\section{Impact on the detectability of TDEs}
\label{fraction}

In the previous section, we argued that the K-H instability can lead to the dissolution of a significant part of the stream before it comes back to pericentre, which could significantly reduce the luminosity emitted from the associated event. Furthermore, we showed that events involving more massive black holes and/or evolved stars are more sensitive to this effect. In this section, we examine the consequence on the detectability of TDEs produced by the disruption of a $1.4 \msun$ star in different evolutionary stages and by black hole of different masses.

For an event to lead to a substantial flare, the star must be disrupted outside the black hole's Schwarzschild radius $\rs$. Otherwise, it is swallowed whole without significant emission. We investigate the effect of the K-H instability on the detectability of events satisfying this condition. To this aim, we define the probability of such events to occur when the star is in a given evolutionary stage by
\be
f^{\rm stage}_{\rm flaring} = N_{\rm stage} / N_{\rm lifetime},
\ee
where $N_{\rm stage}$ and $N_{\rm lifetime}$ are the number of events occurring during the evolutionary stage and the whole stellar lifetime respectively. The possibility of an event to be affected by the K-H instability is only included in $N_{\rm stage}$. These numbers are obtained by
\be
N_{\rm stage}=\int_{t_{\rm start}}^{t_{\rm end}} \dot{N} \, \chiswa \, \chikh \, \rm dt,
\label{nstage}
\ee
\be
N_{\rm lifetime}=\int_{0}^{t_{\rm lt}} \dot{N} \, \chiswa \, \, \rm dt,
\label{nlt}
\ee
where $t_{\rm start}$ and $t_{\rm end}$ are the starting and ending times of the stage respectively, while $t_{\rm lt}$ is the lifetime of the star. $\dot{N}$ is the disruption rate, which we assumed to scale as $\dot{N} \propto \rt^{1/4}$ following \citet{macleod2012}. $\chiswa$ and $\chikh$ are binary functions given by
\be
\chiswa = \cases{ 0 & \mbox{if}  $\rt \leq \rs$ \cr 1 & \mbox{otherwise} \cr },
\ee
\be
\chikh = \cases{ 0 & \mbox{if} $\fkhp \geq 1$ \cr 1 & \mbox{otherwise} \cr },
\ee
which are respectively zero if the star is swallowed whole and if the stream is affected by the K-H instability according to the criterion defined in Section \ref{interactions}.

This probability is shown in Fig. \ref{fig6} as a function of the black hole mass for different evolutionary stages. The left panel does not take into account the K-H instability, artificially fixing $\chikh=1$ in equation \eqref{nstage}. It reproduces figure 14 (right panel) of \citet{macleod2012}. For $\mh\gtrsim 10^8\msun$, the evolutionary stage of most disrupted stars switches from main sequence stars to giant stars. This is because $\rt<\rs$ for main sequence stars above this mass. Instead, the right panel of Fig. \ref{fig6} includes the effect of the K-H instability. The grey zone indicates affected events in a galaxy with $\eta=5$. For $\mh\gtrsim 10^8\msun$, giant stars as previously become more likely to be tidally disrupted than main sequence stars. However, as giant stars are more sensitive to the K-H instability, all the events are affected by the K-H instability for $\mh\gtrsim 10^9\msun$, which could significantly hamper their detection. The blue hatched region indicates the zone swept by the boundary of the grey area for values of $\eta$ varying from 1 to 1000 from right to left. For $\eta \gtrsim 10$, even the events involving main sequence stars are affected by the K-H instability.

\section{Discussion and conclusion}
\label{conclusion}

The interaction between the debris stream produced by TDEs and the background gas of quiescent galaxies has often been neglected, on the basis of their large difference in density. In this paper, we have investigated this interaction for the bound part of the stream, involved in the flaring activity of these events. Through an analytical argument, we have demonstrated that the K-H instability can affect the debris, especially for disruptions involving an evolved star and/or a massive black hole. In this case, a substantial fraction of the tidal stream can dissolve into the background gas before it reaches pericentre, likely leading to a flare dimmer than previously expected.

In order to model the stream, we have used the analytical model of \citet{lodato2009} for the specific energy distribution within the stream, which assumes that the star is unperturbed until it reaches pericentre. Actually, numerical simulations have shown that the stellar structure is perturbed at pericentre \citep{lodato2009,guillochon2013}. However, this effect can be easily accounted for within the same analytical model, by applying a homologous expansion of the unperturbed model by a factor $\sim 2$ \citep{lodato2009}, which makes the energy distribution very close to the one obtained through simulations. This leads to a stream slightly more resistant to the K-H instability but does not affect our main conclusions.

Another assumption of the model is a total disruption of the star by the black hole. However, simulations have shown that a surviving core can remain after the disruption \citep{guillochon2013}, which keeps following the initial stellar orbit. This likely causes the marginally bound part of the stream to contain less mass than expected from \citet{lodato2009}. The debris returning to pericentre at late times would therefore be even more sensitive to the K-H instability.

We also note that we have neglected the effects of magnetic fields in the stream. Such effects might prevent the dissolution of the stream by the K-H instability \citep{mccourt2015}. We can address this issue analytically by adding a term $\ame \simeq B^2_{\parallel} /(\rhoe \he)$ to the left-hand side of condition \eqref{cond}, where $B_{\parallel}$ is the component of the magnetic field parallel to the stream \citep[section 106]{chandrasekhar1961}. This term is significant only if $\ame \gtrsim \are$, which translates to $B_{\parallel} \gtrsim \rho^{1/2}_{\rm g,e} \ve \simeq 3 \, \mathrm{G} \, (\eta/1)^{1/2} (\rstar/12 \rsun)^{-1} (\mh/10^8 \msun)^{3/10}$, evaluating the right-hand side at $\rt$ and for a stellar radius corresponding to profile RG1. Such values of $B_{\parallel}$ correspond to typical surface magnetic fields for main sequence stars. They probably exceed typical surface magnetic fields in red giants, estimated from magnetic flux conservation in the expansion phase. For example, a expansion by a factor of 10 implies a magnetic field reduced by a factor of 100. In the stream, the critical value for $B_{\parallel}$ is unlikely to be reached for several reasons. Firstly, they require that the star is exactly stretched in the direction of its magnetic field, which is unlikely since the magnetic field orientation is random. Secondly, the magnetic field in the inner region of a star is likely tangled and not ordered in the same direction. In this configuration, magnetic reconnection may also occur in the stretching process, lowering the total magnetic field. In addition, although flux conservation imposes that the magnetic field in the direction of the stream is conserved since the stream stays thin, magnetic diffusion could lead to a decrease of this component as the stream orbits around the black hole. A caveat in these arguments is the ill-known value of the magnetic field strength inside giant stars. Nevertheless, we consider it unlikely that magnetic fields would prevent the K-H instability from developing. However, a definite answer would require to follow the evolution of the stellar magnetic field during the disruption and the fallback of the debris.

Finally, our calculations are made in an ambient medium at rest although an inward velocity of the gas environment could diminish the effect of the K-H instability. We have tested the dependence of our results on this assumption by introducing an radial velocity of the gas, which results in a lower relative velocity in equations \eqref{cond} and \eqref{tkh}. We find that our main conclusions remain unchanged for an infall velocity up to the Keplerian velocity, thus confirming the solidity of our analysis.

The main implication of this study is that any TDEs involving black holes with masses $\gtrsim 10^8 \msun$ might be difficult to detect, a conclusion largely independent of our scaling for the background gas density with black hole mass. This was already known for main sequence stars, which are swallowed whole for this range of masses \citep{macleod2012}. Here, we show that this is also the case for giant stars, which have their debris stream dissolved into the background gas through the K-H instability.

\section*{Acknowledgments}

C.B. is grateful to Giovanni Dipierro for hosting him during the completion of this work.

\label{lastpage}

\bibliography{biblio}

\end{document}